Attilio. Sacripanti
ENEA Physics Methods and New Materials Dpt.
Biomechanics of Sport   Rome   University of Tor Vergata ( Medicine Faculty)

Abstract
Movement and Man at the end of the Random Walks

In this paper, it is presented the well known aspect of non linearity of internal human body structures.
Similarity on the basis of the Fractional Brownian Motion from the static ones, as the geometrical fractals like the Intestine and placenta linings, Airways in lungs , Arterial system in kidneys and so on. To the kinematics ones, as the temporal fractals like Heart beat sequences, Electroencephalograms, Respiratory tidal volumes, DNA sequences mapping and so on.
But this internal well known knowledge is astonishing extendible from the internal, through the "Brownian" basis of the muscular contraction, to the macroscopic external human movements, like the orthostatic equilibrium, gait, running training, judo contest and soccer, basketball, football or water polo games, by means of the ubiquitous continuous presence; the Brownian Dynamics.

Movement and man at the end of the Random Walks

1 Introduction –fractals in Human body Physiology
2 Inside the Body
2.1 Fractal dimension, Self -Similarity, Self- Affinity.
2.2 Fractals as Geometrical Self Organization.
2.3 Gauss and Pareto Inverse Power Law.
2.4 Random Walk and its limits.
2.5 Continuum limit of Fractional Random Walks.
2.6 Time series : some example of internal body answers.
2.7 Myosin II, Brownian Ratchet and Muscular Contraction.
2.8 On the boundary
3 Outside the body
3.1 Fluctuation of Surface Body Temperature
3.2 Human Balance : Centre Of Pressure  (COP)  vs  Centre of Mass (CM)
3.3 Multifractals in Human Gait Normal and diseases.
4 From Usual Movement to Sport Movement
4.1 Multifractals in Running Training
4.2 Situation Sports
4.3 Dual Sport With Contact
4.4 Active Brownian Motion
4.5 Team Sports.
5 Conclusions.
6 Bibliography

Attilio Sacripanti

Movement and Man at the end of Random Walks

1 Introduction
Fractals in Human Body Physiology

Starting from the Mandelbrot work (1977) fractal is most often associated with irregular geometric objects that display self similarity.
In an idealized model, this property holds on all scale, but in real world this is not all truth.
In the human body a number of complex anatomic structures display fractal like geometry, that in our special language we will call them "Static", related to the irregular geometric definite form of the structure.
Many of these self-similar structure serve at least at one fundamental physiologic function: rapid and efficient transport over complex spatially distributed structures or organs.
Fractal geometry also appears to underlie important aspects of heart mechanical functions. and brain structure.
Really in the human body a variety of other organs systems contain fractal-like structure that ease information dissemination ( nervous system) or, nutrient absorption ( bowel) , all that shows the ubiquity of "Static" fractals in the human body.
The fractal concept can be found out, in the human body, not only as "static" like the irregular geometric structures, but also in that we will call in our special agreement "Kinematics" form of fractals.
They are connected to body's complex processes that generate irregular fluctuations across multiple time scales
A easy understanding of this behaviour can be obtained by plotting their fluctuations at different time resolutions.

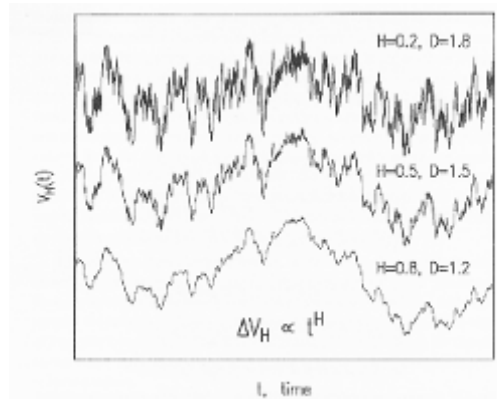

2 Inside the Body

2.1 Fractal Dimension and Self Similarity and Self affinity

Since the fractals occupy an intermediate position between standard geometric subject with integer dimensions, they can be conveniently characterized by their fractal dimension
Then an important parameter describing fractal geometrical structures is the fractal dimension which generalizes the usual integer valued topological dimension., for each structure it was empirically found that the total length L varies as a power of the length scale l

$$L(l) \propto l^{1-D} \quad (1)$$

The parameter D is called fractal dimension of the curve, another way to write the equation (1) is

$$L(l) \propto N(l) \cdot l \Rightarrow N(l) \propto l^{-D} \quad (2)$$

Where N(l) is the minimum number of boxes with side l needed to cover the fractal curve.
There are several generalization of the fractal dimension but it is possible find them in every fractal book.
However data are usually a sequence of real numbers ( time series ) and in this case we may have not information about detail of the problem that we call Kinematics.
Assuming that the time series is determined by an unknown deterministic dynamics it os still possible under general conditions, to reconstruct its phase space and analyze the system "Kinematics".
This is the foundation of the non linear time series analysis. In effect we can obtain from the dynamical equation of the system an apparent random time series that reveals simple structure of the dynamics when embedded in $R^2$.
Fractal dimensions, as seen, can be introduced in various distinct way, each emphasizing a different geometric aspect of the pattern.
An important property of a fractal is its self-similar nature. In other words if we magnify some fragment of such pattern ( both static geometrical or kinematics temporal ) we would see precisely the same structure reproduced on a new scale.
Moreover if the miniature copy may be distorted in other way, for example, somewhat skewed , in this case we call this quasi-self-similarity , self- affinity.
A favourite example of self similarity in many textbook is the trajectory of the Brownian Motion of a particle. Although this trajectory is extremely irregular, it displays some remarkable invariance, if we take a small region containing part of the trajectory, and enlarge it the result would look very similar to the original whole trajectory.
In the next figure we can see two example of self-similarity the first geometric the second time related.

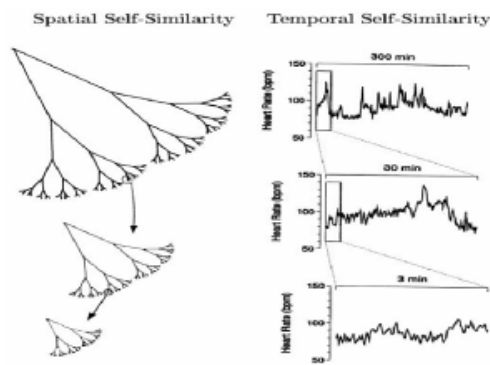

2.2 Fractal as Geometrical Self Organization

We speak of self – organization or, more exactly, of self organized behaviour if every part of the complex system acts in a well defined way on the given internal order or configuration.
To show in a rigorous mathematical terms the self organization , it is possible on the track of Haken, Liupanov and others to develop a very general theory applicable to a large class of different systems comprising not only sociological systems, but also physical, chemical , and biological systems.
If the internal organizing force is obeying themselves equations of motion.
In the general case of a chemical systems we show that the solution is not linear, if $q_1$ is a function of the concentration of the chemical and the "effect" are $q_2$ it is possible to write for a self-organized chemical system:

$$\begin{cases} \dot{q}_1 = -\gamma_1 q_1 - aq_1 q_2 \\ \dot{q}_2 = -\gamma_2 q_2 - bq_1^2 \end{cases} \qquad 3)$$

$$q_2(t) \approx \frac{b}{\gamma_2} q_1^2$$

$$\dot{q}_1 = -\gamma_1 q_1 - \frac{ab}{\gamma_2} q_1^3 \qquad 4)$$

These equations (3) show in very simple way that the self similar structures in the human body could got by nonlinear (Brownian) self - organized chemical reactions (4).
If the system possesses a macroscopically infinitesimal time scale, so that during any dt on that scale all of the reaction channels fire many more times than once yet none of the propensity functions change appreciably, we can approximate the discrete Markov process by a continuous Markov process defined by the Chemical Langevin (SDE) Equation

$$X_i(t+dt) = X_i(t) + \sum_{j=1}^{n} v_j a_j X(t) dt + \sum_{j=1}^{n} v_j a_j^{\frac{1}{2}} X(t) N_j(t)(dt)^{\frac{1}{2}} \qquad (5)$$

where $N_1 \ldots N_m$ are M temporally uncorrelated, statistically independent normal variables with mean 0 and variance 1
In the specific structures of the human body like, coronary artery tree or Purkjne cells in cerebellum or in the human heart these Fractal geometrical Self-Organization comes more easily from the corresponding Fokker Plank equation (5):

$$\dot{f}(q_u, q_s) = \left\{ \frac{\partial}{\partial q_u}(\gamma_u q_u + aq_u q_s) + \frac{\partial}{\partial q_s}(\gamma_s q_s - bq_u^2) \right\} f(q_u, q_s) + + \frac{1}{2}\left( K_u \frac{\partial^2}{\partial q_u^2} + K_s \frac{\partial^2}{\partial q_s^2} \right) f(q_u, q_s) \quad (6)$$

In several cases like chemical reaction dynamics, a Fokker Planck equation is more easy to obtain than the corresponding Langevin equation.
This short speech, show very clearly, that Random Walk or is continuous limit the Brownian motion are at the basis of the fractal in human physiology.
In the next figures there are shown two well known example of physiological self-organization the coronary artery tree, and the Purkinje cells in cerebellum.

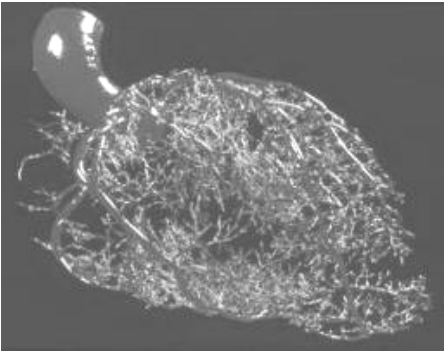
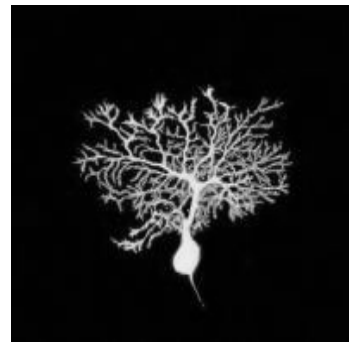

## 2.3 Gauss and Pareto Inverse Power Law

Chance in physics is connected to the concept of probability, many process in real life are random and their dynamic evolution is very difficult to understand from a deterministic point of view.
Probability theory born to explain the outcome of games, one of the simplest games is tossing a coin where one can find head or tail.
Normally the number of proof are determined and finite ( binomial distribution) but more known is the limiting case to the infinite proof number, this analysis was performed by Gauss.
In definite form we, today, can define the world of Gauss " simple" as scientific world view, in this ( linear) theory the output is proportional to the input, the algebra is additive, the presence of simple rules yield simple result of the problem, these results are stable, the phenomenon are predictable, their result stable and the final distribution obviously Gaussian.
In mathematical form the system evolution is defined by the following equation:

$$\frac{dX}{dt} = \mu X + \xi(t) \qquad (7)$$

In this Langevin like equation we can see that the fluctuation is simply additive and the bell curve that define the Gauss distribution is the well known simple inverse curve of the normal distribution:

$$P(x) \propto e^{-\left[\frac{\mu}{D}x^2\right]} \qquad (8)$$

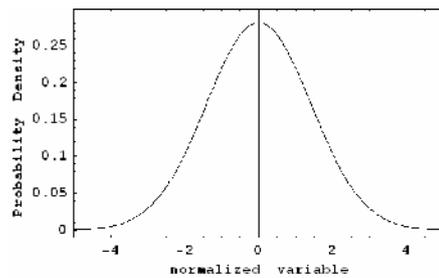

But a more complex scientific vision is connected to a not well known great Italian scientist, except outside the social and economic world, Pareto.
By his law, it is possible to describe a more complex scientific vision of the world, this world is non linear, in which small changes may produce divergence in the solutions.
It is a multiplicative world, in which simple rules yield complex results, the processes are unstable the predictability is limited and the description of the phenomena is both qualitative and quantitative, the Pareto distribution it is also an inverse power law distribution.
In mathematical form the system evolution is defined by the following equation:

$$\frac{dX}{dt} = \mu X + \xi(t) X \qquad (9)$$

In this more complex world fluctuation is multiplicative and the Pareto inverse power law distribution satisfies the following form:

$$P(x) \propto \frac{1}{|x|^{1+\frac{\mu}{D}}} \qquad (10)$$

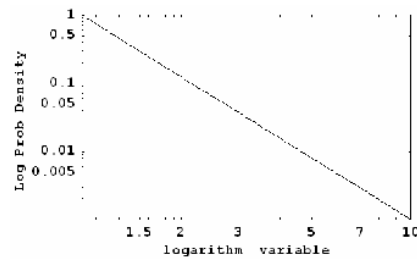

## 2.4 Random Walk and its limits

If we consider a stochastic process going on time, for example the motion of a particle which is randomly hopping backward and forward; this example is known in the scientific world as "Random Walk" . If we would know the probability that after n hopping the particle will be in a position m , it could be common to connect this probability in mathematical discrete form and to write:

$$P(m; n+l) = w(m, m-l)P(m-l; n) + w(m, m+l)P(m+l; n) \quad (11)$$

This discrete probability equation has two very important limits in his continuum form. The limit of the random walk with infinitesimal, independent steps is called Brownian Motion. The first form is the limit that explains more carefully the dynamic of a single particle in time and its mathematical form is known as Langevin Equation from the French Scientist that proposes it in 1905

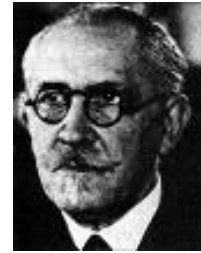

$$\frac{dv(t)}{dt} = -\mu v(t) + \xi(t)$$

(12)

In which the first term after the equal is the dissipation suffered by the particle and the second is the stochastic fluctuation applied on it and, if Gaussian type, this dissipation will be zero in mean over time.

The second form, is the limit that take in account more the point of view of the global probabilistic aspect of the random process analyzed in phase space , and its mathematical form is well known as Fokker Planck Equation from the two German Scientists that propose it in 1930.

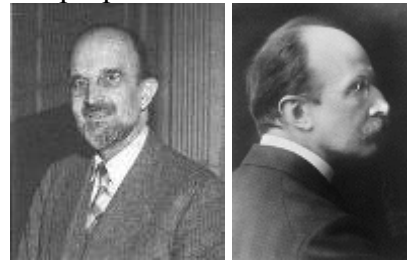

$$\frac{\partial P(v,t)}{\partial t} = -\frac{\partial}{\partial v}\left[-\mu v P(v,t)\right] + \frac{\partial^2}{\partial v^2} D P(v,t)$$

(13)

In the next figure (showing in it also the self similarity property of this random process) we can see one example of two dimensional Random Walk or Brownian Motion of a particle, that as before described, it is view in two different conceptual way, by the two previous limit equation.

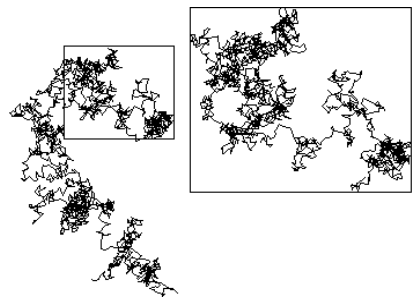

## 2.5 Continuum limit of fractional random walks

An interesting way to approach complex systems derives also from a special view of the Random walks, it is based to incorporate this complexity in it introducing memory in the random walks through fractional differences, this generalization has the same significant limits previously introduced.

If we see more carefully at the dynamic aspect of the process it is possible to write a generalized fractional Langevin equation and to introduce the Fractional Brownian Motion.

In mathematical form it is possible to write:

$$D_t^\alpha [X(t)] - \frac{X(0)}{\Gamma(1-\alpha)} t^{-\alpha} = \xi(t) \qquad (14)$$

In which the first term is a fractional derivative, the second is connected to the initial condition of the process, and the third is always the random force acting on the particle.

In this case is important to know the mean square displacement of the particle:

$$\left\langle [X(t) - X(0)]^2 \right\rangle = \frac{\langle \xi^2 \rangle}{(2\alpha - 1)\Gamma(\alpha)^2} t^{2\alpha - 1} \propto t^{2H} \qquad (15)$$

From this expression it is possible to understand, that we are in presence of an anomalous diffusion process, identified by the H parameter usually called Hurst parameter, in particular this parameter is time independent, and it describes the fractional Brownian motion with anti-correlated samples for 0<H<1/2 and with correlated samples for ½<H<1 if H is = to ½ we can speak of Brownian Motion. If in general H could be a function of time.. In recent time this important extension is called multifractional Brownian motion..

This important generalization come from certain situation occurring either in the field of turbulence ( Frisch 1999) or from Biomechanics ( Collins and De Luca 1994) where there are the needs of a more flexible model necessary both: to control locally the dependence structure and to allow the path regularly to vary with time.

*Pure Brownian motion: next step is uncorrelated with previous step H=0.5*

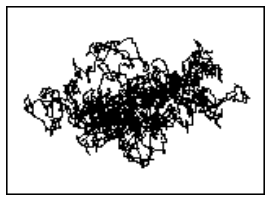

*Persistent Fractional Brownian motion: each step is positively correlated with previous step H<0.5*

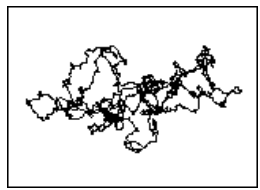

*Anti-Persistent Fractional Brownian motion: each step is negatively correlated with previous step H> 0.5*

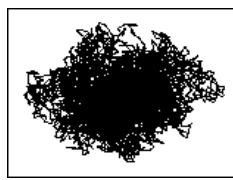

2.6 Time series : some example of internal body answers.

Many organs inside the human body could be controlled or analyzed by instrumentation which grip electric time series signals as answers of the specific organ.
For example Brain imaging data may generally show fractal characteristics - self-similarity, 1/f-like spectral properties ( like Pareto inverse power law ).
Normally Self-similar or scale-invariant time series like EEG, ECG, have 1/f-like power spectrums with these accepted classification
– if a = 0, noise is *white*
– if a = 2, noise is *brown*   (random walk )         $P_f \approx |f^{-\alpha}|$   (16)
– if a = 3, noise is *black*   (Nile floods)
– if 0 < a < 2, noise is *pink*  (J. S. Bach)
Fractional Brownian Motion has covariance parameterised by Hurst exponent 0 < H < 1
The Hurst exponent, the spectral exponent a, and the fractal (Hausdorf) dimension FD, are simply related:  2H+1 = a   or   2-H = FD  For example classical Brownian motion has a = 2, H = 0.5 and FD = 1.5  as easily it is seen in the next figure

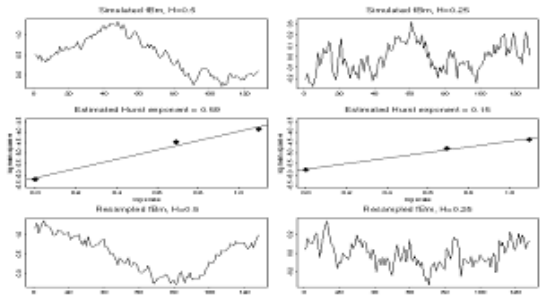

Wavelets are the natural basis for analysis and synthesis of fractal processes in the human body, they are used in the brain image analysis, but the same method could be applied to the heart analysis. In the first two figures we can see, as example of diseases in white blood cells and circulatory dynamics. In the next four, respectively, it is possible to see:
One example of Severe Congestive Heart Failure, on signal of a Healthy Heart, again a Severe Congestive Heart Failure et the last example is the signal of a Cardiac Arrhythmia, Atrial Fibrillation

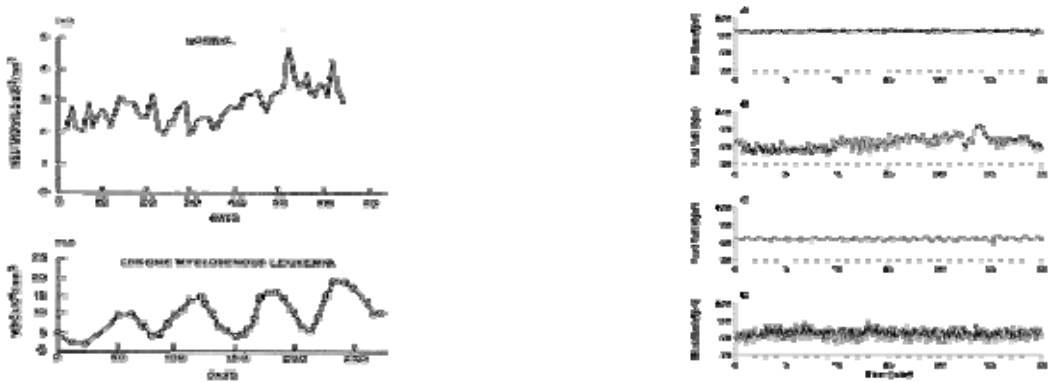

Another interesting properties of these signals is that  Fractal Complexity  degrades with Disease.

In the next example it is possible to see a degradation of a signal from complexity to a simplified signal index of disease.

Healthy Dynamics: Multiscale Fractal Variability

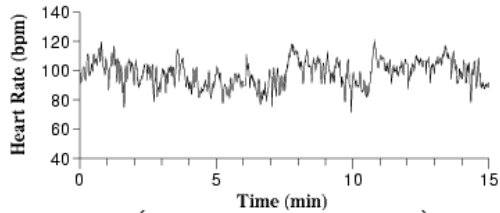

*Two Patterns of Pathologic Breakdown*

Single Scale Periodicity  Uncorrelated Randomness

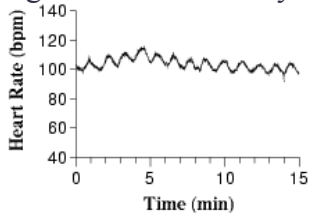 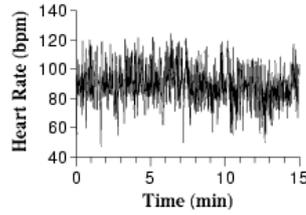

Another hypothesis is connected to the time irreversibility of the signal; time irreversibility is greatest for healthy physiologic dynamics, which have the highest adaptability. Time irreversibility decreases with aging and disease in the first heart rate in the next   Human Respiration: Loss of Long-Range  (Fractal) Correlations with Age

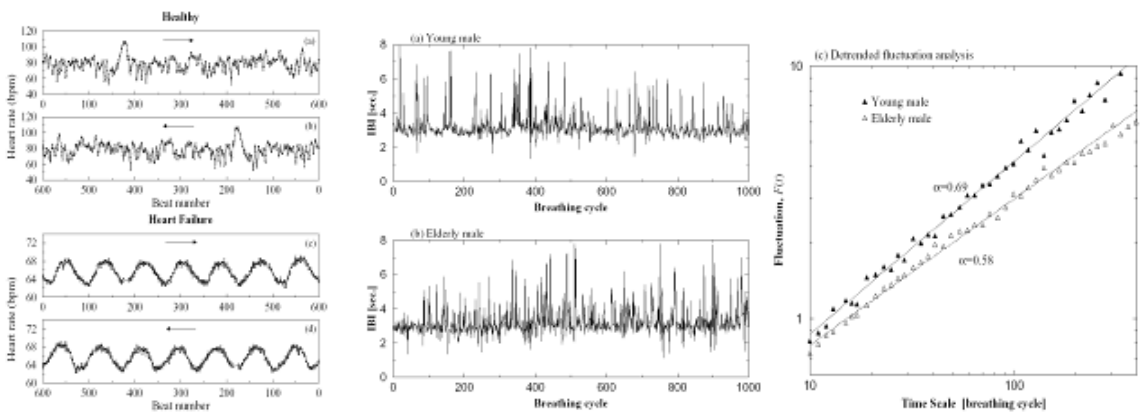

2.7  Myosin (II)  Brownian Ratchet and the Muscular Contraction

Now it experimentally well known that the model of Brownian Ratchet , give us a good explanation of the non processive motor , the Myosin II , using the thermal fluctuation and also the energy stored in the ATP  structure, can  move long the Actin  filament
Till now two models are in competitions the Huxley and Simmons  power stroke ( lever arm) model and the Brownian ratchet.
The problem is to compute the real Myosin translation  to evaluate the correctness of the model.
The translation caused by the pivoting of the lever arm would be about 5 nm. New technologies for manipulating a single actin filaments allow to test the lever arm model  but the displacement varied considerably some report have shown myosin displacement of about 5nm which is consistent with the lever arm model.

However others have shown that if myosin is oriented correctly relative to actin filament axis as in muscle the value increases to 10-15 nm out of model predictions.

For the ratchet model one problem is that the scale of the motion is smaller than the Brownian motion of microneedles, because the average amplitude is between 30-40 nm.

Recently Yanagida et others 2000, has shown, manipulating a single myosin head and measuring the displacements with a scanning probe. This assay allowed measurement of individual displacements of single myosin head with high resolution. The data showed that a myosin (II) head mowed along an actin filament with single mechanical steps of 5.5 nm; groups of two to five rapid steps in succession often can produce displacements of 11 to 30 nm.

Similar sub-steps observed are constant in size with the repeat actin monomers (5.5 nm), independent of force, and because some subsets are backward, it is more likely that a Myosin head may step along the Actin monomer repeat by biased Brownian Motion.

In the next figure we can see for the Myosin II and V, both the models:

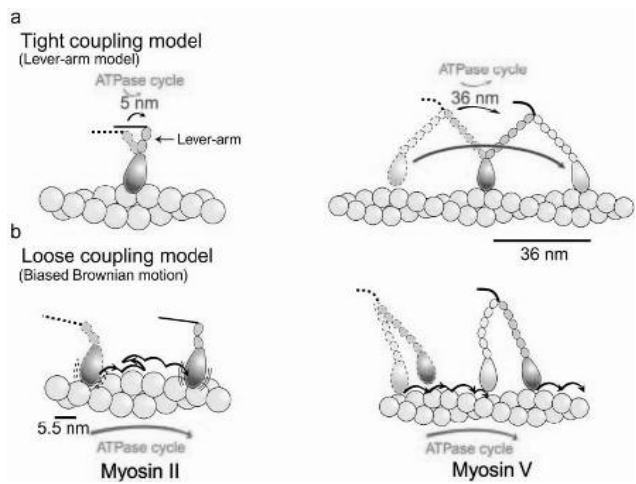

2.8 On the Boundary

In this years a very interesting model as Random Walk Model of the Human Skin Permeation was presented by Frash in 2002 in this model the skin normally made by heterogeneous different material in different layer, was presented first as homogeneous membrane: made by the same material

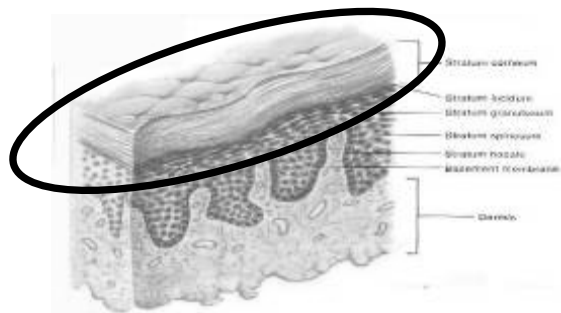

Normally Effective path length can be defined as the thickness of a homogeneous membrane having identical permeation properties as the skin, and the Effective diffusivity is a diffusivity of a homogeneous membrane having identical permeation properties as the body's skin.

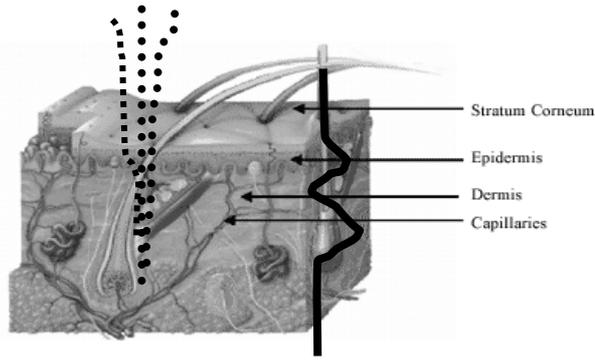

The next figure shows the interesting results of this random walk model of the human skin
Random walk simulations for mass penetration through SC with log Kcor_lip=0
Dcor/Dlip = 1.0/0.01

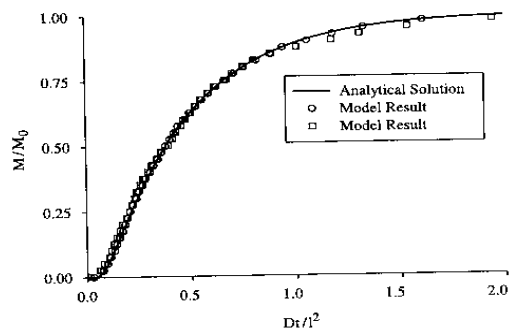

## 3  Outside the Body

As there is shown the geometric fractals and temporal fractals are widely common inside the human body, but if we remember that, also at microscopic level, Brownian motion is ubiquitous like in DNA, molecular motors for the seven myosin family, axon transmission command etc. this means that also the basis of the muscular contraction are Brownian, then by scheletric muscles we can go outside the human body , to find other interesting presence of the Brownian motion, and astonishingly there are many others random process or situations in which the Brownian motion normal, fractal and also multi fractals are present.
In the following paragraphs there are show some interesting examples.

3.1 Fluctuation of surface body temperature.

If the surface body temperature fluctuation are analyzed for example by a thermocamera we can see that the equation describing this topological situation in steady state condition, in which the $\zeta(t)$ is the random fluctuation of the incident absorbed radiation. And $\Delta T(t)$ is the surface body difference of temperature depending from the time and the position

$$c_v \frac{d\Delta T(t)}{dt} + h\Delta T(t) = \zeta(t)$$  (17) Langevin like equation

$$\left\langle (E - \overline{E})^2 \right\rangle = kT^2 c_v$$  (18) Fluctuation in energy

$$\left\langle (\Delta T)^2 \right\rangle = \frac{kT^2}{c_v}$$  (19). Fluctuation in temperature.

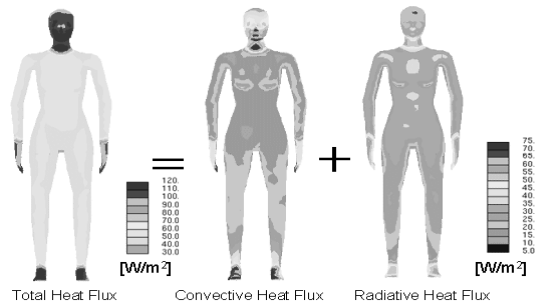

The previous equation shows that the surface temperature fluctuation of the human body is Brownian.
But the topology of the surface temperature taken by thermocamera easy shows that the superficial temperature is a function not only of time but also of position.
In the next frame the last consideration is very clear

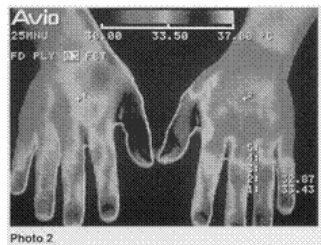

If the body makes a motion or works in not controlled condition, then the equation considered as a whole could take this form in his free evolution captured by thermocamera:

$$c_v \frac{d^2 \Delta T(t)}{dt^2} + \left(h + \frac{dc_v}{dt}\right)\frac{d\Delta T(t)}{dt} - \frac{d\zeta(t)}{dt} = 0 \qquad (20)$$

Really speaking the situation is more complex because the specific heat $c_v$ is not constant, but it is a very complicate function at least of time, space, food, fat, and body dimensions; some works in this field to identify the real shape of this function, where made by Sacripanti and co-workers about ten years ago (1995-1997).
In fact it is a mistake or a first approximation, to identify a constant number as specific heat for the human body that is, in reality, a complex engine with continuous production and dispersion of energy (the metabolic heat).
In this case the function $c_v$ better called "body's Thermal Inertia" is not constant but a complex function with shape as lennard-Jones potential.

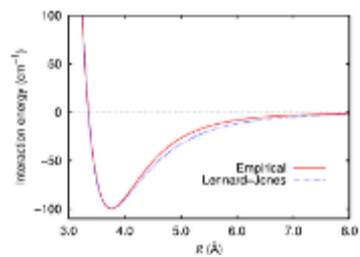

Also h the thermal coefficient is not really constant but is dependence is very more complex because for work like normal activity or sport movement in not controlled thermal condition, it must satisfy the following experimental Sacripanti's relationship:

$$S\sigma\varepsilon\left(\frac{T_s^4 - T_a^4}{t - t_0}\right) + 0.6n\frac{kS \operatorname{Re}^{0.8} \operatorname{Pr}^{0.33}}{l}\frac{T_i - T_a}{t - t_0} + \left\{e^{-\frac{4S}{lh}\frac{T_s - T_b}{T_b - T_a}}\right\}$$

$$\left\{\left[0.132\varepsilon_h \frac{4S^2 k \operatorname{Re}^{0.8} \operatorname{Pr}^{0.33}}{hl^2}\frac{(T_s - T_a)^{1.2}}{T_a^{0.2}(t - t_0)}\right] + \left[0.16(1-\varepsilon_h)\frac{4S^2 D\lambda \operatorname{Re}^{0.8} \operatorname{Sc}^{0.33}}{Rl^2 h}\left(\frac{M_s e_s}{T_s} - \frac{M_a e_a}{T_a}\right)\frac{(T_{vs} - T_{va})^{1.2}}{T_{va}^{0.2}(t - t_0)}\right]\right\}$$

$$\left\{e^{-(0.2\varepsilon^2 + 0.5\varepsilon - 0.7)\frac{\lambda P - \Sigma}{\lambda P}} - 1\right\} = \frac{dO_2}{dt}$$

In the next figures we can see some thermograms of a judo technique taken by a thermocamera in a very pioneering work of the author and co–workers (1989) in which the thermal emission was connected to the oxygen consumption by means of the previous Sacripanti's relation.

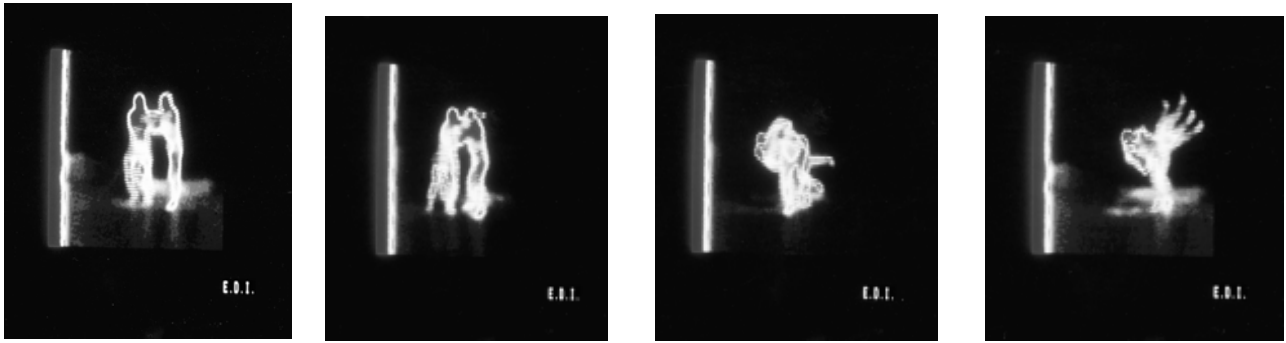

3.2 Human balance Centre Of Pressure ( COP ) vs. Centre of Mass ( CM)

In static equilibrium the CM ( centre of Mass) and the COP (centre of pressure) would lie on the same vertical line COP would coincide with the projection of the CM on the ground.
This can be illustrated in Biomechanics using a simple model, the inverted pendulum, Winter 1998, Pedotti 1987 .for the anterior posterior balance.
The pendulum rotates around the ankle joint which we take as origin of the Cartesian system if we denote as F the force acting on foot by force plate at the point (-ζ, η) which is the COP.
The system is described in Newtonian approximation by the equations:

$$m\ddot{y} = F_y$$
$$m\ddot{z} = F_z - mg \qquad (22)$$
$$I\ddot{\alpha} = \eta F_z + \zeta F_y - mgL\cos\alpha$$

The component $F_z$ is the same force as is obtained from the readings of the force transducers. For a small deviation around the vertical z axis, we may replace cosα by y/L and in the first approximation we may also set $F_z = mg$ then the last equation will be :

$$y - \eta \approx \left(\frac{\zeta}{g} + \frac{I}{mgL}\right)\ddot{y} \qquad (23)$$

After some easy manipulation and putting the equation in term of the angle π/2- α= θ we obtain:

$$\ddot{\alpha} - \left(\frac{mgL}{I}\right)\alpha = 0 \qquad (24)$$

This equation of course describes an unstable situation, the inverted pendulum topples over. However this classical procedure do not explain the Random Walks characteristics of quiet standing coordinates of the COP they can be explained by the equation of Hastings & Sugihara 1993 that combines random walk with a friction term like a Langevin equation:

$$dx(t) = -rx(t)dt + dB(t) \qquad (25)$$

Here dB is the uncorrelated noise with zero mean

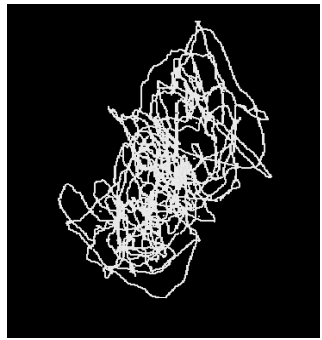

Posturogram- Random Walk of the COP coordinates.

3.3 Multifractals in Human Gait

Walking is a very complex voluntary activity, the typical pattern shown by the stride interval time series suggest particular neuromuscular mechanisms that can be mathematical modelled.
The fractal nature of the stride time series of human was incorporated into a dynamical model by Hausdorff using a stochastic model that was later extended by Askhenazi et others so as to describe the analysis of the gait dynamics during aging.
The model was essentially a random walk on a Markov or short range correlated chain, where each node is a neural that fires an action potential with a particular intensity when interested by the random walker.
This mechanism generates a fractal process with a multifractals aspects, in which the Holder time dependent exponent depends parametrically on the range of the random walker's step size.
The multifractal gait analysis is also used to study the fractal dynamics of body motion for patients with special aging problems or diseases, like Parkinson or post-stroke hemiplegic.
In the next figures we can see.
The variation of the time dependent Holder exponent , with the walker step size for free pace and metronome pace at different speed.
The different time series produced by free pace and metronome pace at different speed.
The relative acceleration signals and the related fractal values in post stroke and Parkinson patients.

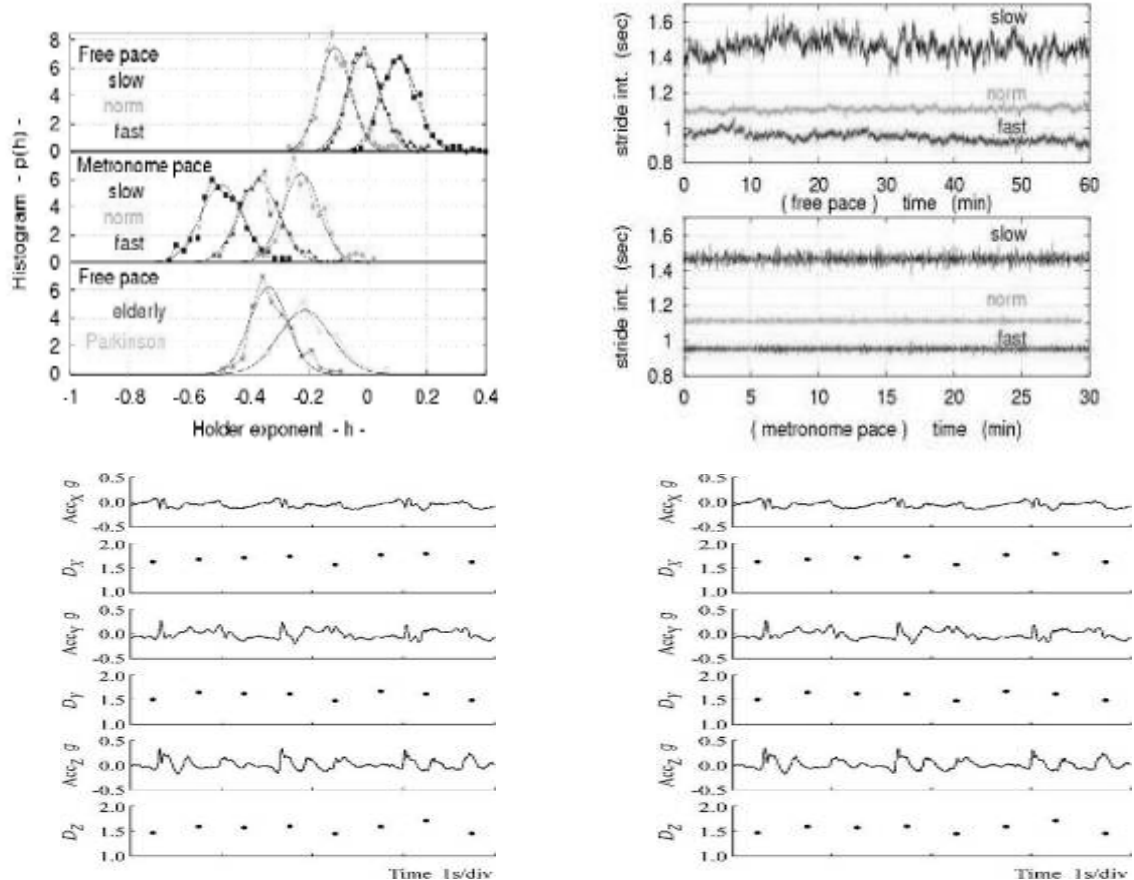

## 4 From Usual Movement to Sport Movement

4.1 Multifractals in Running training

The fractal nature of the physiological signals: heart and respiratory frequencies and oxygen uptake in long distance runs was compared and analyzed by Billat et others 2001-2002-2004.
Today middle and long distance running are characterized by speed variability.
The statistics show that if there are considered the last three world record on a middle or long distance running, it can be observed that velocity varies of 5%.
In races the variation of the velocity and the choice of the optimal speed , obviously involve a complex interplay between physiological and psychological factors .
Multifractal analysis is used in biomechanics of running for classifying signals which exhibit a rough behaviour .
This behaviour is quantified calculating the holder exponent using multifractal analysis.
The following equations give the formula of the exponent:

$$H(t_0) = \liminf_{t \to t_0} \frac{\log|f(t) - P_m(t-t_0)|}{\log|t-t_0|} \qquad (26)$$

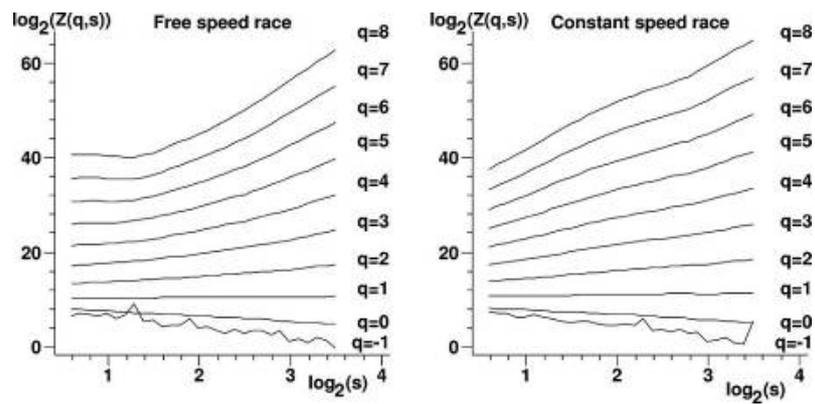
Free and constant heart rate scaling law behaviour

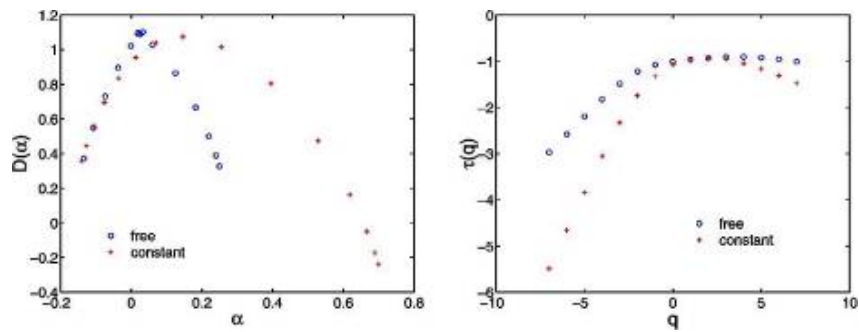
Heart rate spectrum and scaling exponent

4.2 Situation Sport

In the field of Biomechanics of Sport it is interesting to classify the sport , in order to study the athletic performance.
There are many potential classifications, for example in function of performance energy expenditure, but one of the more useful one's is the biomechanical classification that is in function of the most basic movement performed during the performance.
This classification allows to single out the most basic complex movement that must be measured by specific scientific discipline as whole or in step size.
In alternative this most basic movement could be the goal of an expert group like: Sport Physiologist, Neurologist, Biomechanics, engineering, trainers, technicians.
The aim of this classification is to allow to find rightly what kind of specific observational approach must be applied to solve the problem by mechanical or mathematical models both qualitative or quantitative one's.
This classification allows us to group all the sports in four big families.

Cyclic Sports
There are all the sports in which the basic movement is repeated continuously in time like: gait, running, marathon cycling, swimming, etc.

Acyclic Sports
There are all the sports in which the basic movement is applied only once in the performance, like : discus, shot put, hammer throw, pole vault, high jump, long jump, triple jump, ski jump, javelin throw, etc.

Alternate-Cycling Sport
There are all the sports in which two basic movements are applied alternatively in time, like 110 hurts , 400 hurts, steepchase , golf.

Situation Sport
There are all the sports with the presence of the adversary.
These sports, can be divided in two classes (without and with contact) and each class in two sub classes *dual sports* and *team sports.*
The first dual ones are tennis and ping pong, and as team sports we can find: volleyball, beach volley.
The last dual ones, in which athletes can contact together, are: fighting sports: judo, boxing, wrestling, karate, etc. and as team sport, soccer, basketball, football, water polo, hockey, etc.
The situation sports are sport in which it is not possible to find a repeatable motion pattern for each specific game- For each game the motion in it, is a random process, then there are not basic specific movement during the motion, but it is possible to find these repeatable movements .only during the interaction among athletes.
In fact the correct way to analyze such macro phenomena is to study them. in two steps: motion and interaction with basic repeatable movements..
And we can find astonishing that motion for each class of these sports could be associate at one of the previous Brownian Motion that we show.
In fact if we consider for each sport the motion basic pattern of a big number of games from the statistical point of view, like classical Gaussian approach, it appears easily that the motion belongs to the classes of Brownian Motion.

4.3 Dual sport

We take in consideration the couple of athletes as a single system, then the motion of the centre of mass system is definite by a push pull random forces. That in formulas can be express as:

$$\varphi(t) = u \sum_j \delta(t - t_j) \qquad (27)$$

The system is isolated no external forces less the random push-pull forces then the motion equation will be a Langevin like equation, ( Sacripanti 1992 ) and it is possible to write:

$$F = ma = -\mu v + u \sum_j (\pm 1)_j \delta(t - t_j) = F_a + F' \qquad (28)$$

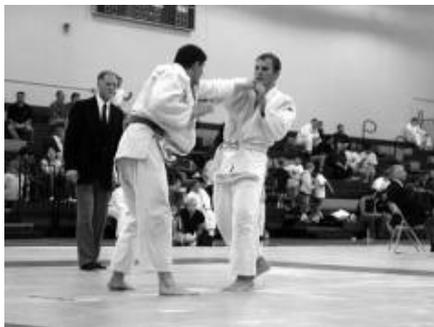

Tacking in account the well known work of Smoluchovski on the Brownian Motion , the " Physical that produce the random evolution of the contest allows us to obtain the basic probability of this Markovian process.

Then for dual sports it is possible to obtain from the transition probability Q the solutions of conditional Probability, which give at infinite time limit the probability to find an athlete between x and x + dx at time t, in mathematical form we can write:

$$Q(k,m) = \frac{1}{2}\delta(m, k-1) + \frac{1}{2}\delta(m, k+1) \quad (29)$$ that give us the solution

$$P(n|m,s) = \frac{s!}{\left(\frac{v+s}{2}\right)!\left(\frac{v-s}{2}\right)!}\left(\frac{1}{2}\right)^s \quad (30)$$

The experimental proof of this model can be founded in some Japanese works, on the world championship of the 1971.
In the next figure we can see the summation of motion patterns of 1,2,7, and 12 contests of judo .it is easy to see that the random fluctuation not have a preferential direction over the time this means that $\langle F' \rangle = 0$ and the motion of the Centre of Mass systems is Brownian.

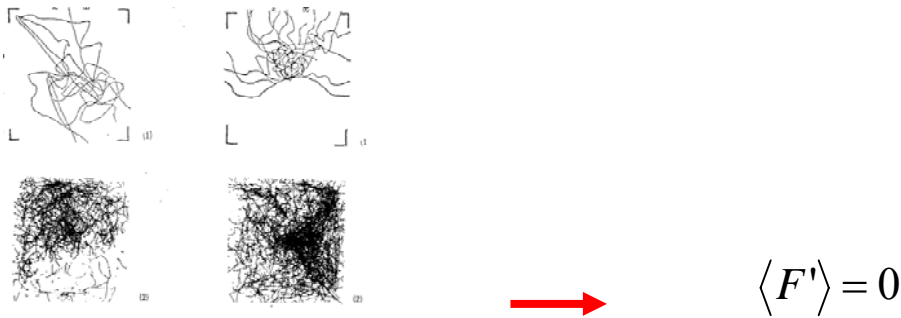

$$\langle F' \rangle = 0$$

4.4 Active Brownian Motion

Motion in team sport is better modelized by the active Brownian motion proposed by Ebeling and Schweitzer tacking in account the oxygen uptake from the particles and its consumption from the environment.
Particle with mass $m$, position $r$, velocity $v$, self-propelling force connected to energy storage depot $e(t)$; velocity dependent friction $\gamma(v)$, external parabolic potential $U(r)$ and noise $F(t)$ could satisfy the following Langevin like equation that could be wrote:
This is the motion system equation for the active Brownian motion

$$m\,\partial_t v = d_2\,e(t)\,v - \gamma(v)\,v - \nabla U(r) + F(t)$$
$$\partial_t r = v \quad (31)$$

From the original work of Ebeling et co-workers it is possible to take in account the energy depot: space-dependent take-up q(r), the internal dissipation c e(t), and also the conversion of internal energy into kinetic energy d2 e(t) v2 then the relative equation is:

$$\partial_t\,e(t) = q(r) - c\,e(t) - d_2\,e(t)\,v^2 \quad (32)$$

Energy depot analysis (for $q(r) = q0$) that are special constant conditions, gives the following result, after some calculation, in term of friction non linear coefficient:

$$\gamma(v) = \gamma_0 - \frac{d_2 q_0}{c + d_2 v^2} \qquad (33)$$

This data can be specialized for human people playing in team games, as it is shown in the following pages.

4.5 Team sport

In the first Sacripanti's model, relative to the dual situation sports, the motion of the centre mass of couple of athletes systems is a classical Brownian Motion, that means there is not a special direction in their motion patters.
In the case of team sport, the situation is completely different, there is, in mean, a special preferential direction in motion pattern and every single athlete is not in stable equilibrium as the couple of athlete's system.
In this case, it is necessary to adopt a different model for the motion, like the active Brownian motion proposed by Ebeling and Schweitzer., to take in account the oxygen uptake from the environment. In this special case it is possible to write for the energy depot variation :

$\frac{dE(t)}{dt} = \dot{V}(t) - \eta(v^2) K v^2 E(t)$  If we take the hypothesis that the energy E(t) is slowly varying, the previous equation can be simplified on the basis of the following considerations:

$\frac{dE(t)}{dt} \approx 0$

$\dot{V}(t) \cong \overline{V}(t)$

Then it is possible to obtain the special value for the energy $E_0(t)$ namely :

$E_0 = \frac{\overline{V}}{\eta K v^2}$  the equation achieves a term $kE_0 v$  as shown by Ebeling, the friction coefficient in this case will be:

$$\gamma_v = \gamma_0 + E_0 \equiv \gamma_0 - \frac{k\overline{V}}{\eta k v_0^2} = \gamma_0 - \frac{\overline{V}}{\eta v_0^2} \qquad (34)$$

considering also the potential interaction against the adversary , collision or avoidance we can present the second Sacripanti model.
On the basis of the Ebeling and Schweitzer model and the Helbing equation, the following Langevin type equation proposed in the Sacripanti's second model accounts of:
motion, oxygen uptake, kinetic energy from uptake and potential mechanical interaction like collision and avoidance manoeuvres:

$$ma = \left(\gamma_0 - \frac{\overline{V}}{\eta v^2}\right) v(r,t) + \frac{m}{t}\left[v^0 e(t) - v_1\right] + \theta k(r_{1,2} - d_{1,2}) N_{1,2} +$$

$$+ A_{1,2} N_{1,2} e^{\left(\frac{r_{1,2} - d_{1,2}}{B}\right)} \left[\lambda_1 - (1-\lambda_1)\frac{1+\cos\varphi_{1,2}}{2}\right] + u\sum_j (\pm 1)\delta(t - t_j) \qquad (35)$$

In compact form it is possible to write

$$ma = -\gamma_v v + F_{acc} + \left[\Sigma F_1 + \Sigma F_2\right] + u\sum_j (\pm 1)\delta(t - t_j) =$$

$$= -\gamma_v v + F_{acc} + \left[\Sigma F_1 + \Sigma F_2\right] + F' \qquad (36)$$

The specific preferred direction in motion patterns of the team sports, could be modellized by the solution model proposed by Erenfest, but with a special modification made by the author it is possible to modellize the basic probability of this Markovian process ( the game) in function of the special attack strategy utilized. In effect for the team sports it is possible to write the transition probability Q in function of the attack strategy α. The α parameter can vary from 1 to 5 , with these meanings :
1= lightning attack ; 2= making deep passes; 3= manoeuvring ; 4= attack by horizontal passes; 5= melina

The solution of the conditional probability P are connected to the limit of mean value in time for finding the athlete between x and x + dx at time t, in formulas:

$$Q(k,m) = \frac{R^\alpha + k}{2R^\alpha}\delta(m,k-1) + \frac{R^\alpha - k}{2R^\alpha}\delta(m,k+1)$$
$$con -1 \leq \alpha \leq 5 \qquad (37)$$
$$\langle m(s) \rangle_{av} = \sum_m mP(n[m,s]) = \left(1 - \frac{1}{R^\alpha}\right)\langle m(s-1) \rangle_{av}$$

In the next four figures it is possible to see that, in spite of the preferential direction present in each motion pattern, from the statistical point of view ( summation of several motion patters from several games) also in team games the random fluctuation not have a preferential direction over the time, this means that $\langle F' \rangle = 0$ and the global motion also in this case is Brownian.

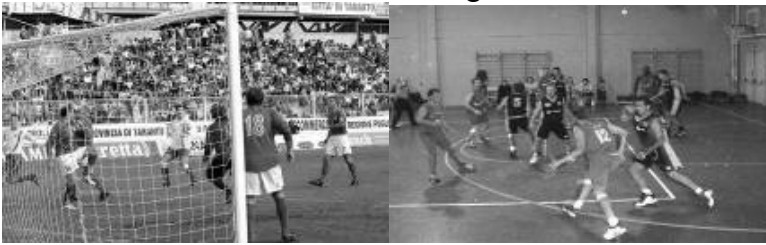

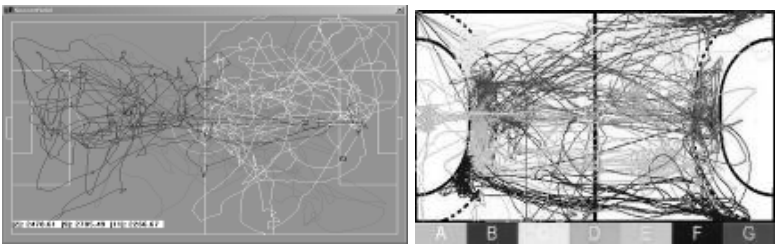

$\langle F' \rangle = 0$

## 5  Conclusions

From this article it is understandable that all the self organizing complex systems, especially the biological ones, like the human body , are better described by non linear evolutions equations that show themselves in their static, kinematics and dynamics forms. ( fractals ).
The only connection among these different aspects is the generalized Brownian Motion in every known formulation : classic, fractional, active and so on.
Its results starting from fractals till to multifractals aspects assure us, at light of our knowledge , that Brownian dynamics is one of the basic modelling of the mathematical alphabets of Life.

# 6 Bibliography

Billat V. et al. Fractal analysis of Speed and Physiological Oscillations in long-and Middle distance Running Effect of training  Int. Juor. Of Computer Science in Sport  Vol 2 (2003) 16-30,
Gaspard P. Brownian motion,dynamical randomness and irreversibility  New Juor. of Physics 7 (2005)77 .
Marques de Sà  J. Fractals in Physiology  Universitade do Porto  2001
Ebeling, Schweitzer  Active Brownian Particles with energy depots modelling animal mobility BioSystems 49 (1999) 17-29
Lakoba et al. Modifications of the Helbing-Molnar-Farkas-Vicsek social force model for pedestrian evolution  Simulation  1 ( 2005) 339-352.
Klimontovich Yu. Nonlinear Brownian motion  Physics- Uspekhi  37 (1994) 737-767
Castellanos-Moreno  Movimiento browniano activo mediante velocidades estocastica  Rivista Mexicana de fisica 49 (2003) 429-438.
Schweitzer  Active Brownian particles with internal energy depot  Traffic and granular flow Springer Berlin 2000   161-172.
Schweitzer Active motion of Brownian particles  Stochastic process in Physics Chemistry and Biology  Springer Berlin 2000  97-107.
Ebeling   Nonlinear Brownian motion mean square displacement   Condensed matter physics   7 (2004) 539-550
Morters & Peres  Brownian Motion  Draft version   May 2006
Rudnick  Lectures on Random walks   2001
Ebeling et al. Klimontovich's contributions to the kinetic theory of nonlinear Brownian motion and new developments  journal of physics 11 (2005) 89-98
Nelson  Dynamical Theories of brownian motion  Princeton University second edition  2001
Faris   Lectures on stochastic processes   2001.
Bruce&Gordon  Better Motion prediction for people tracking  Paper  2002
Montemerlo & Wittaker  conditional particles filter for simultaneous mobile robot localization and people tracking  IEEE  (ICRA)  2002
Silver-Torn  Investigation of lower limb tissue perfusion during loading  Journal of rehabilitation research and development  35  (2005) 597-608
Sekine et al  Fractal dynamics of body motion in patients with Parkinson's disease Journal of neural engineering 1 (2004)  8-15
Akay et al   Fractal dynamics of body motion in post stroke hemiplegic patients during walking Journal of neural engineering 1 (2004) 111-116
Yanagita & Iwane  A large step for Myosin  pnas  97 (2000) 9357-9359
Verdier   Rheological properties of living materials from cells to tissues  Journal of theoretical medicine  5 (2003) 67-91.
Frank  Stochastic feedback, non linear families of Markov process and nonlinear  Fokker-Planck equations  Physica A  331 (2004) 391-408.
Quian   The mathematical   theory of molecular motor movement and chemomechanical energy transduction  journal of mathematical chemistry  27 (2000) 219-234-
Borg  Review of nonlinear methods and modelling 1  Biosignal projects  2002.
Scafetta  et al. Holder exponent spectra for human gait  Physica A 328 (2003) 561-583
Goldberger et al   Fractals dynamics in physiology: alterations with disease and aging   pnas   99 (2002) supplement 1
Coeurjolly   Identification of multifractional Brownian motion  Paper  2004
Benassi et al. identifying the multifractional function of a Gaussian process   Statistics and probability letters 39 (1998) 337-345.

Lauk et al  Human balance out of equilibrium nonequilibriunm statistical mechanics in posture control  Physical review letters 80 (1998) 413-416.
Maciejewski et. Alt .*The impact of autonomus robots on crowd behavior*  Colorado state university 2003
Belair et alt. *Behavioral dynamics for pedestrian*   final draft  2003
Sacripanti et Al.   Serie - Valutazione del costo energetico degli sport  dicombattimentoin remote sensing: Progress report N° 2  " *Intercomparazione operativa*" ENEA-RT- INN / 90/06
Sacripanti.   Serie - Valutazione del costo energetico degli sport  di combattimento in remote sensing: Progress report N° 3 " *Teoria biomeccanica della competizione*".
ENEA-RT- INN/  90 / 07.
Sacripanti et Al.   Serie -Valutazione del costo energetico degli sport di combattimentoin remote sensing: Progress reports N°4/5. " *Stima dell'effetto schermo del judogi* "ENEA-RT- INN / 91 / 50.
Sacripanti et Al.   Serie - Valutazione del costo energetico degli sport  di combattimento in  remote sensing:Progress report N° 6"*Valutazione dell'effetto schermo del judogi* "ENEA-RT-INN /91/ 51
Sacripanti, Dal Monte, Rossi. Serie - Valutazione del costo energetico degli sport  di combattimento in remote sensing: Progress report N° 7. "*Man-environment , heat - exchange equations a new thermodynamic approach* " Eighth meeting of the European Society of Biomechanics Rome 92 ENEA-RT- INN/ 92/ 08.
 Sacripanti et Al.   Serie - Valutazione del costo energetico degli sport  di combattimento in remote sensing: Progress report N° 8. " New trends in the judo and  wrestling biomechanics research " Eighth meeting of the European Society of Biomechanics Rome 92  ENEA-RT- INN/ 92/ 09
Sacripanti et Al.   Serie - Valutazione del costo energetico degli sport  di combattimento in remote sensing: Progress report N° 9   "*Physical work and thermal emission* "
X International Symposium of Biomechanics in Sport - Italy 1992   ENEA-RT- INN/ 92/ 1
Sacripanti, Dal Monte, Rossi.    Serie - Valutazione del costo energetico degli sport  di combattimento in remote sensing: Progress report N° 10  " *Man-Environment heat exchange equation, evolution and improvement*"  ENEA-RT- INN/ 93/ 32
 Sacripanti et Al.   Serie - Valutazione del costo energetico degli sport di combattimento in remote sensing: Progress report N° 11 " *Apparato sperimentale e procedura esecutiva* "ENEA-RT- INN/ 94/  15
Sacripanti   *Biomeccanica del Judo*    ed mediterranee   1989
Sacripanti   *Biomeccanica degli  Sport*    ed il vascello 2004
Mc Garry  *Sport competitions as a dynamical self organizing system*   Jouof sport science 20 2002
 Pers  et alt *Errors and mistakes in automated player tracking*    computer video workshop 2003   .
Bloomfield et alt.*Temporal pattern analysis and his applicability in soccer*   The hidden structure of interaction  IOS Press 2005
 Yamamoto et al  On the fractal nature of heart rate variability in humans: effects of vagal blockade  The American physiology society  1995
 Tempaku  Random Brownian motion regulates the quantity of human immunodeficiency  virus type1 (HIV-1) attachment and infection of target cell Journal of health science 51 (2005) 237-241.
Deligneres et al.  Time interval production in tapping and oscillatory motion Human movement science  23 (2004) 87-103
Burggren & Monticino  Assessing physiological complexity  the journal of experimental biology 208 (2005) 3221-3232.
Westfreid at al   Multifractal analysis of heartbeat time series in human race   Applied and computational harmonic analysis  18 (2005)  329-335.
Frasch, A Random Walk Model of Skin Permeation, Risk Analysis.Vol.22,No.2, 2002
A. Schätzlein, G. Cevec. Non-uniform cellular packing of the stratum corneum and permeability barrier function of intact skin: a high-resolution confocal  laser scanning microscopy study using highly deformable vesicles (Transfersomes). Br.J.Dermatol.138, 1996